\documentclass[aps,pra,twocolumn]{revtex4-2}

\usepackage{graphicx}
\usepackage{float}
\usepackage{physics}
\usepackage[hidelinks]{hyperref}
\usepackage[capitalise,noabbrev]{cleveref}
\usepackage{multirow}

\begin{document}

\newcommand{\goo}{g_{11}}
\newcommand{\gii}{g_{ii}}
\newcommand{\aii}{a_{ii}}
\newcommand{\aoo}{a_{11}}
\newcommand{\att}{a_{22}}
\newcommand{\aot}{a_{12}}
\newcommand{\gtt}{g_{22}}
\newcommand{\got}{g_{12}}
\newcommand{\Elhy}{E_{\mathrm{LHY}}}
\newcommand{\varEmf}{\mathcal{E}_{\mathrm{MF}}}
\newcommand{\varElhy}{\mathcal{E}_{\mathrm{LHY}}}


\title{Trapped Imbalanced Quantum Droplets}

\author{T. A. Flynn}
\email{t.flynn@ncl.ac.uk}
\author{N. A. Keepfer}
\author{N. G. Parker}
\author{T. P. Billam}
\affiliation{Joint Quantum Centre (JQC) Durham--Newcastle, School of Mathematics, Statistics and Physics, Newcastle University, Newcastle upon Tyne, NE1 7RU, United Kingdom}



\date{\today}

\begin{abstract}
A two-component quantum droplet is an attractive mixture of ultracold bosons
stabilised against collapse by quantum fluctuations. Commonly, two-component
quantum droplets are studied within a balanced mixture. However, the mixture
can be imbalanced resulting in a lower energy but less stably bound droplet, or
even a droplet submerged in a gas. This work focuses on the experimentally
relevant question: how are imbalanced droplets modified by harmonic trap
potentials? Droplet ground states and breathing modes are analysed across the
two-dimensional parameter space of imbalance and trap strength. The robustness
of the droplet imbalance is also studied by releasing the droplet from the
trap, demonstrating that this can lead to the creation of free-space,
imbalanced droplets. 
\end{abstract}


\maketitle

\section{INTRODUCTION \label{sec:intro}}

Quantum gases have developed into a rich platform to study a variety of physics
from analogues of condensed matter and many-body systems,
\cite{bloch2005,bloch2008,bloch2012}, to simulators of cosmological processes
\cite{roger2016,eckel2018,proukakis2017}. Much of the theoretical and
experimental results of these studies are dominated by Mean-Field (MF)
contributions.  At ultracold temperatures quantum mechanical effects are
pronounced, enabling the study of Beyond-Mean-Field (BMF) contributions, i.e.,
quantum fluctuations. Two-component quantum droplets are one such quantum gas
system in which quantum fluctuations are highly significant
\cite{petrov2015,petrov2018}.

Quantum droplets can be formed in ultracold mixtures of atomic Bose gases, in
which the interactions between the two species are tuned to be dominantly
attractive. This highlights another advantage of quantum gases: there is
precise control of interactions. Two-body interactions --- characterised by the
scattering length, $a_s$ --- can be tuned via a Feshbach resonance
\cite{courteille1998,inouye1998,chin2010}. Taking into account only MF physics,
these attractive mixtures would unstable to collapse. The collapse of the cloud
leads to an increased density and consequently an increased contribution of the
BMF corrections. Quantum fluctuations lead to an effective repulsion between
the atoms, The repulsion of quantum fluctuations --- described to first order
by the Lee-Huang-Yang (LHY) correction \cite{lee1957} --- balances the
attractive collapse forming a self-bound, dilute liquid droplet
\cite{petrov2015, petrov2018}. Therefore, quantum droplets are an
experimentally observable state of matter in which quantum fluctuations not
only contribute, but are integral. It should be noted that the arrest of
collapse from BMF corrections does not carry across to the single-component
Bose gas, which has been experimentally demonstrated to be unstable to collapse
under attractive two-body atomic interactions \cite{roberts2001,donley2001}.

As indicated above, quantum gases are a platform for probing physics from areas
that appear disparate. One field that has benefited from a close connection
with quantum gases is fluid dynamics. Many quantum gases exhibit superfluidity
which has been used as an analogue to the dynamics of classical fluids such as
vortex dynamics \cite{barenghi2001} or turbulence
\cite{barenghi2014,barenghi2023}. Quantum droplets are a further extension of
this tradition as they can be used to probe liquid properties such as surface
tension \cite{petrov2015,ancilotto2018,2ciko2021} and incompressibility
\cite{baillie2017,ferioli2019}. 

Two-component quantum droplets have been experimentally observed in both
homonuclear $^{39}$K \cite{cabrera2018,semeghini2018,cheiney2018,ferioli2019}
and heteronuclear, $^{41}$K-$^{87}$Rb and $^{23}$Na-$^{87}$Rb
\cite{derrico2019,guo2021}, mixtures. The benefit of using homonuclear mixtures
is the precise control over the population numbers of each component.
Homonuclear mixtures are made of atoms prepared in different hyperfine states.
Experiments begin with all atoms in one component; a radio frequency pulse is
then used to controllably transition a proportion of the atoms to the second
component. This control allows for probing one of the predictions of
two-component droplets: density balancing.

The original droplet prediction of Ref.~\cite{petrov2015} argues that a density
balance is preserved during the droplet formation. This density balance is due
to an energetic favourability for the two component densities to maintain a
fixed ratio $n_2/n_1 = \mathrm{const.}$ where $n_i$ is the number density of
the $i$th component. The majority of theoretical studies of two-component
quantum droplets assume this density balance. However, recent works
\cite{mithun2020, tengstrand2022, flynn2022, ancilotto2023,vallesmuns2023} have
explored the impact of removing this assumption and the properties of such
imbalanced droplets.

Imbalanced droplets fall into two main regimes \cite{petrov2015,
flynn2022}: (1) bound, imbalanced droplets, in which there is a population
imbalance in the droplet core; (2) saturated, imbalanced droplets,
corresponding to a droplet core that is saturated with majority component
atoms, with any further majority component surrounding the droplet as an
unbound gas. 

Ref.~\cite{flynn2022} studied the ground states and breathing modes of
imbalanced quantum droplets in free space. A saturated droplet will lose any
unbound atoms in free space; in a trap, the surrounding gas will be retained as
it is energetically favourable for the gas to sit at the trap minimum. The key
focus of this paper is to investigate how the ground states and breathing modes
are modified with the application of an isotropic harmonic trap. These
investigations are motivated by the feasibility of imbalanced quantum droplets
to be created and probed experimentally. Additionally, this work explores the
stability of these droplets when released into free space.

This work starts by defining the theory used to model quantum droplets in
\cref{sec:mod}. This model is first implemented in \cref{sec:drop_gs_trapped}
to explore how the imbalanced droplet ground states are modified by isotropic,
harmonic trapping potentials. \cref{sec:breath_trap} looks at propagating these
ground states in time, subject to an initial perturbation, to analyse the
droplet breathing modes, both for varying trap strength, and size of imbalance.
\cref{sec:tof} investigates the stability of imbalanced droplets under an
instantaneous removal of the trapping potential, as this is a method widely
used in droplet experiments. Finally, the main conclusions and future work are
discussed in \cref{sec:disc}.

\section{THE MODEL \label{sec:mod}}
A zero-temperature mixture of two weakly-interacting, dilute, homonuclear Bose
gases can be described by the energy functional \cite{ho1996,petrov2015}
\begin{equation}
\begin{split}
  E = \int&\bigg[\frac{\hbar^2}{2m}|\grad{\Psi_1}|^2 + \frac{\hbar^2}{2m}|\grad{\Psi_2}|^2 + V_1|\Psi_1|^2 + V_2|\Psi_2|^2 \\
          &+ \varEmf + \varElhy \bigg]\dd^3\mathbf{r},
\end{split}
\label{eq:e_func}
\end{equation}
in which $m$ is the atomic mass of both components and $V_i$ is the trapping
potential applied to the $i$th-component. The first two terms of
\cref{eq:e_func} are the kinetic energy contributions, whilst $\varEmf$ is
the MF energy density term given by, 
\begin{equation*}
  \varEmf = \frac{2\pi\hbar^2\aoo}{m}|\Psi_1|^4 + \frac{2\pi\hbar^2\att}{m}|\Psi_2|^4 + \frac{4\pi\hbar^2\aot}{m}|\Psi_1|^2|\Psi_2|^2,
\end{equation*}
where $a_{ii}$ and $\aot$ are the intra- and inter-species scattering lengths.
The final term, $\varElhy$, is the energy density of the LHY correction which,
to first-order, describes the effects of quantum fluctuations on the condensate
\cite{lee1957}. For a homonuclear bosonic mixture the LHY correction takes the
analytic form \cite{petrov2015}
\begin{equation}
  \varElhy = \frac{256\sqrt{\pi}\hbar^2}{15m}\left(\aoo|\Psi_1|^2 + \att|\Psi_2|^2\right)^{5/2}.
  \label{eq:e_lhy}
\end{equation}
The LHY energy density does not depend on $\aot$ due to the assumption that the
mixtures lies at the critical point of attractive instability, i.e., $\aot^2 =
\aoo\att$, removing the issue of complex contributions resulting from an
unstable phonon mode \cite{petrov2015,qi2020,xiong2022}. It should be noted
that this approximation is made only in the derivation of \cref{eq:e_lhy}, and
does not imply any parameter choice in later sections.

The energy functional in \cref{eq:e_func} can be minimised via the variational
relation $i\hbar(\pdv*{\Psi_i}{t}) = \fdv*{E}{\Psi_i}$, giving the equal-mass,
coupled extended GP equations \cite{petrov2015} 
\begin{align}
  \begin{split}
  i\hbar\pdv{\Psi_1}{t} = &\bigg[-\frac{\hbar^2}{2m}\laplacian + V_1 + \frac{4\pi\hbar^2}{m}\left(\aoo|\Psi_1|^2 + \aot|\Psi_2|^2\right) \\
  &+ \frac{128\sqrt{\pi}\hbar^2\aoo}{3m}\left(\aoo|\Psi_1|^2 + \att |\Psi_2|^2\right)^{3/2}\bigg]\Psi_1, \\
  i\hbar\pdv{\Psi_2}{t} = &\bigg[-\frac{\hbar^2}{2m}\laplacian + V_2 + \frac{4\pi\hbar^2}{m}\left(\att|\Psi_2|^2 + \aot|\Psi_1|^2\right) \\
  &+ \frac{128\sqrt{\pi}\hbar^2\att}{3m}\left(\aoo|\Psi_1|^2 + \att |\Psi_2|^2\right)^{3/2}\bigg]\Psi_2.
    \label{eq:dimmed_gpes}
  \end{split}
\end{align}
The dimensional scalings $\mathbf{r} = \xi\mathbf{\tilde{r}}$,
$t=\tau\tilde{t}$ and $\Psi_i = \rho_i^{1/2}\tilde{\Psi}_i$ result in the
dimensionless, equal-mass coupled extended GP equations,
\begin{align}
  \begin{split}
    i\pdv{\Psi_1}{t} = &\bigg[-\frac{1}{2}\laplacian + V_1 + |\Psi_1|^2 + \eta|\Psi_2|^2 \\
    &+ \alpha\left(|\Psi_1|^2 + \beta|\Psi_2|^2\right)^{3/2}\bigg]\Psi_1, \\
    i\pdv{\Psi_2}{t} = &\bigg[-\frac{1}{2}\laplacian + V_2 + \beta|\Psi_2|^2 + \eta\beta|\Psi_1|^2 \\
    &+ \alpha\beta^2\left(|\Psi_1|^2 + \beta|\Psi_2|^2\right)^{3/2}\bigg]\Psi_2.
    \label{eq:dim_gpes}
  \end{split}
\end{align}
in which all tildes have been neglected and the dimensionless parameters are 
\begin{equation*}
  \eta = \frac{\aot}{\sqrt{\aoo\att}},
\end{equation*}
\begin{equation*}
  \alpha = \frac{32}{3}\left[\frac{2}{3\pi}\frac{|\delta a|\aoo^{5/2}n_1^{(0)}}{\sqrt{\aoo} + \sqrt{\att}}\right]^{1/2}, \quad \beta = \left(\frac{\att}{\aoo}\right)^{1/2},
\end{equation*}
with dimensional parameters
\begin{align*}
  \xi = \sqrt{\frac{3}{8\pi}\frac{(\sqrt{\aoo} + \sqrt{\att})}{|\delta a|\sqrt{\aoo} n_1^{(0)}}}, & & \tau = \frac{3m}{8\pi\hbar}\frac{(\sqrt{\aoo} + \sqrt{\att})}{|\delta a| \sqrt{\aoo} n_1^{(0)}}, \\ \rho_1 = \frac{2}{3}\frac{|\delta a|n_1^{(0)}}{\sqrt{\aoo}(\sqrt{\aoo} + \sqrt{\att})}, & & \rho_2 = \frac{2}{3}\frac{|\delta a|n_1^{(0)}}{\sqrt{\att}(\sqrt{\aoo} + \sqrt{\att})}, \\
\end{align*}
where $\delta a = \aot + \sqrt{\aoo\att}$ and $n_1^{(0)}$ is the equilibrium
density of component-1 for the balanced mixture \cite{petrov2015}. The
expression of the equilibrium density is calculated in a homogeneous infinite
system under the criterion of a vanishing pressure --- i.e., the droplet in
equilibrium with the vacuum --- and takes the form \cite{petrov2015}
\begin{equation*}
n_1^{(0)} = \frac{25\pi}{1024} \frac{(\aot + \sqrt{\aoo\att})^2}{\aoo^{3/2}\att(\sqrt{\aoo} + \sqrt{\att})^5}.
\end{equation*}
The density scalings $\rho_i$ correspond to rescaled normalisation constants,
$\tilde{N_i} = N_i/(\rho_i\xi^3)$, in which $N_i$ is the population of the
$i$th-component wavefunction. By breaking the assumption of density-locking, it is
possible to imbalance the population numbers such that $N_2/N_1 \neq
\sqrt{\aoo/\att}$.  

The trapping potentials are non-dimensionalised by $V_i =
\left(m\xi^2/\tau^2\right)\tilde{V}_i$. This work only considers isotropic
harmonic trapping with equal traps applied to each component,
i.e., $V_i = V = \frac{1}{2}m\omega_r^2r^2$. Under non-dimensionalisation this
becomes $\tilde{V} = \frac{1}{2}\tilde{\omega}_r^2\tilde{r}^2$, in which
$\tilde{\omega}_r^2 = \left(\tau\xi^2m/\hbar\right)\omega_r^2$. In subsequent
sections the dimensionless population numbers and trapping potentials are
presented without tildes, as only dimensionless parameters are used hereafter.

The results in this paper will be contrasted with the density-locked model,
used widely in modelling quantum droplet experiments
\cite{cabrera2018,cheiney2018,ferioli2019}. The density-locked model assumes a
constant density ratio, $n_2/n_1 = \sqrt{\aoo/\att}$, such that the two
component wavefunctions can be expressed in terms of a single wavefunction,
$\Psi_i = \sqrt{n_i}\phi$, neglecting any out-of-phase motion between the
components \cite{petrov2015,bienaime2016,capellaro2018}. Under these
assumptions, Equations (\ref{eq:dimmed_gpes}) can be non-dimensionalised and
reduced to a single equation,
\begin{equation*}
i\pdv{\phi}{t} = \left[-\frac{1}{2}\laplacian - 3|\phi|^2 + \frac{5}{2}|\phi|^3\right]\phi,
\end{equation*}
with the system described by a single parameter, an effective atom number,
$\tilde{N}$, given by \cite{petrov2015}
\begin{equation}
\tilde{N} = \left(\frac{\sqrt{\att}}{n_{1}^{(0)}(\sqrt{\aoo} + \sqrt{\att})}\right)\frac{N}{\xi^3},
\label{eq:eff_num}
\end{equation}
in which $N$ here is the total atom number $N = N_1 + N_2$. Within this work,
balanced and imbalanced droplets are both modelled by Equations
(\ref{eq:dim_gpes}), though it should be noted that for a balanced droplet the
dimensionless parameters $(N_1, N_2,\alpha,\beta,\eta)$ can be recast to
$\tilde{N}$. In the density-locked model a given set of scattering lengths,
$a_{ii}$ and $\aot$, correspond to a fixed population number ratio, $N_2/N_1 =
\sqrt{\aoo/\att}$. 

\section{GROUND STATES \label{sec:drop_gs_trapped}}

How the density of a spherically-symmetric balanced droplet varies with harmonic
trap frequency has been studied in Ref.~\cite{hui2020}. The trap frequency can
be considered low if there is no significant change from the free-space droplet
density, whereas a higher frequency trap eventually leads to the flat-top
density of large droplets being lost. Furthermore, in free space the negative
chemical potential, $-\mu$, is described as the particle emission threshold of
the droplet, however this description breaks down in a trap \cite{hui2020}. It
can therefore be argued that in the trap-dominated regime the idea of a droplet
begins to be less defined, and the mixture transitions to a trapped gas.
Ref.~\cite{hui2020} approximates the transition to the trap-dominated regime as
the point in which the potential energy at the droplet surface becomes
comparable to the binding energy of the droplet, resulting in $\omega_{r}^{(c)}
\sim (4\pi/3\tilde{N})^{1/3} $ where $\tilde{N}$ is the effective atom number
of the density-locked model in \cref{eq:eff_num} \cite{petrov2015}.

One main assumption of this work is spherical-symmetry. Density is assumed to
be a function of radius only, reducing the computational problem to an
effective 1D system with the kinetic term becoming $\laplacian\Psi_i
\rightarrow [\pdv*[2]{(r\Psi_i)}{r}]/r$. Another assumption of this work is
balanced intraspecies scattering lengths ($\aoo = \att \implies \beta = 1$).
Thus, the only possibly difference between components is from an imposed
population number imbalance of $N_1 = N_2 + \delta N_1$. To find ground states,
Equations (\ref{eq:dim_gpes}) are evaluated numerically in imaginary time until
the energy of the mixture is adequately converged \footnote{To check for ground
state convergence, the energy difference between two successive time steps,
$\epsilon_{\mathrm{diff}}$, is used. The tolerance for convergence is
$\epsilon_{\mathrm{diff}}\lesssim 10^{-8}$}. The numerical scheme is a
$4^{\mathrm{th}}$-order Runge-Kutta method, using a $2^{\mathrm{nd}}$-order
centred finite-difference scheme for the spatial derivatives. Neumann boundary
conditions ($\pdv*{\Psi_i}{r} = 0$) are applied at $r=0$ and at $r=L_r$, where
$L_r$ is the radial computational box size. Note that for all simulations
presented in this work, $\omega_r^{(c)}\approx 0.186$.

\begin{figure}
  \begin{tabular}{cc}
    \includegraphics[width=0.25\textwidth]{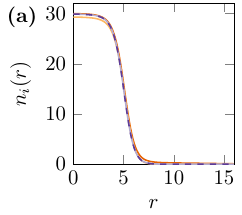} &
    \includegraphics[width=0.25\textwidth]{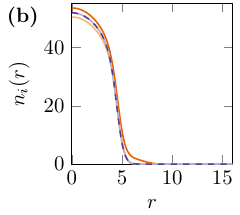}
  \end{tabular} \\
  \includegraphics[width=0.5\textwidth]{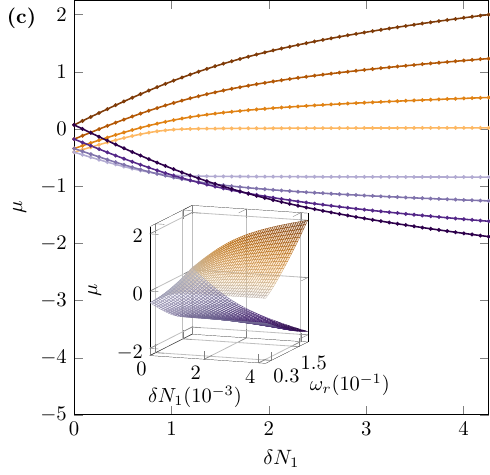}
  \caption{Balanced and imbalanced quantum droplets in isotropic harmonic traps
  (with, $N_2 \approx 17027$, $\alpha\approx0.00657$ and $\eta\approx-1.11$).
(a) Droplet ground state density profiles, of size $\tilde{N} \approx 649$ with
balanced components, $\delta N_1 = 0$, (the light and dark purple dashed lines)
and an imbalance of $\delta N_1 \approx 8513$ (majority component --- dark
orange, minority component --- light orange), in a trap of frequency $\omega_r
\approx 0.0442$. (b) Equivalent balanced and imbalanced droplet to (a), but
with a trapping frequency of $\omega_r \approx 0.353$. (c) Majority (orange)
and minority (purple) chemical potentials across the 2D parameter
space of imbalance and trap frequency --- i.e., $(\delta N_1, \omega_r)$ --- for
the fixed droplet size considered in (a) and (b), at trap frequencies of
$\omega_r \approx \{0.00441,0.0662,0.128,0.190\}$, with $0 \leq \delta N_1
\lesssim 4257$. The inset shows the surfaces for the majority (orange) and
minority (purple) component chemical potentials across the 2D parameter space
for which the curves in (c) are 1D slices of set $\omega_r$ values. Note that
the two surfaces are equal at $\delta N_1=0$ imbalance, but diverge for
increasing imbalance.} 
  \label{fig:gs_trap}
\end{figure}

\cref{fig:gs_trap}(a) shows balanced (purple) and imbalanced (orange) droplet
density profiles in a trap with frequency $\omega_r\approx0.0442$.
\cref{fig:gs_trap}(b) presents an example of the imbalanced atoms forming a
significant gas density around the surface of the droplet in a trap with higher
frequency $\omega_r \approx 0.353$. \cref{fig:gs_trap}(a) shows that the
imbalanced and balanced droplets have comparable density profiles, whereas
\cref{fig:gs_trap}(b) shows a more considerable deviation between
the balanced and imbalanced droplets, in a higher frequency trap. The central
density splitting becomes more pronounced in the higher frequency trap, showing
a more suppressed minority-component central density.

The divergence of the two chemical potentials with increasing imbalance is a
key observation of Ref \cite{flynn2022}. 
For increasing trap frequency the chemical
potential of a balanced droplet increases, eventually becoming positive as the
density of the mixture significantly deviates from the free-space droplet
density. \cref{fig:gs_trap}(c) presents chemical potential data, defined by a
2D parameter space of imbalance and trap frequency $(\delta N_1,\omega_r)$, 
where $\omega_r \approx \{0.00441,0.0662,0.128,0.190\}$, i.e., showing trap
frequencies up to approximately $\omega_r^{(c)}$.

\cref{fig:gs_trap}(c) shows that the two-component chemical potentials of
balanced droplets are equal and increase with trap frequency, in agreement with
Ref.~\cite{hui2020}. Beyond the balanced case, the lowest frequency trap
($\omega_r \approx 0.00441$) shows similar behaviour to the free-space chemical
potentials presented in Ref \cite{flynn2022}, in that the chemical potentials
appear to reach a saturation limit, though there will be effects from the
unbound gas such that these curves are approximate to the saturated limit. For
the higher trap frequencies of $\omega_r \approx \{0.0663,0.128,0.190\}$, the
two chemical potentials diverge with the majority component chemical potential
becoming large and positive, whilst the minority component chemical potential
becomes large and negative. The diverging chemical potentials represent a clear
distinction between balanced and imbalanced trapped droplets, and an excerpt of
the 2D parameter space is included in the inset of \cref{fig:gs_trap}(c), with
the majority and minority chemical potentials plotted as orange and purple
surfaces, respectively, demonstrating the chemical potentials diverging for
increased trap strength and imbalance. 


Adding harmonic traps to both balanced and imbalanced droplets causes the
flat-topped density profile to eventually be lost with increasing trap
frequency. One key difference between balanced and imbalanced droplets is that
the trap causes any unbound atoms to form a trapped gas at the droplet
surface. For balanced droplets the chemical potential increases with trap
frequency until eventually becoming positive. Whereas, for imbalanced droplets
it is always possible for the minority component chemical potential to be made
negative in isotropic harmonic traps by tuning the imbalance. One way to
understand the effect that this squeezed external gas cloud has on the droplet
is to analyse the breathing modes of the droplets.

\section{BREATHING MODES \label{sec:breath_trap}}

To initiate the breathing mode dynamics of trapped droplets, a perturbation is
made by imprinting a harmonic potential of the form $e^{i\epsilon r^2}$, where
$\epsilon$ is small (here $\epsilon = 10^{-5}$) onto the minority component
ground state wavefunction \cite{pit2016,stringari1996}. This perturbed ground
state is then propagated in real time.

The breathing modes of balanced droplets in free space have two regimes,
self-evaporative and non-self-evaporative \cite{petrov2015,fort2021}. In the
self-evaporative regime the breathing mode is unstable because the mode
frequency exceeds the particle emission threshold, $-\mu$. Hence, the droplet
will emit atoms to lower its energy, corresponding to a decaying sinusoidal
oscillation with a frequency that asymptotes to the particle emission
threshold, $-\mu$. In the non-self-evaporative regime, the breathing mode
frequency does not exceed the particle emission threshold therefore the mode is
stable and non-decaying. Additionally, the frequency of the balanced droplet
breathing mode varies with droplet size only \cite{petrov2015}. 

Breathing modes in imbalanced droplets are instead dominated by unstable regions.
For both self-evaporative and non-self-evaporative droplets, an imbalance implies
an unstable, decaying breathing mode except for small imbalances in the
non-self-evaporative regime \cite{flynn2022}. 

In this analysis, the focus will be on decaying breathing modes (i.e.,
excluding non-self-evaporative droplets that are either balanced or have
sufficiently small imbalances). Trapped balanced droplets exhibiting
non-decaying breathing modes have already been studied in Ref \cite{hui2020}.
To examine the breathing modes of trapped, imbalanced droplets, this section
will take one example droplet size, $\tilde{N} \approx 649$ (as in
\cref{fig:gs_trap}). Different droplet sizes yield qualitatively the same
behaviour, with only a difference in mode frequency. By fixing droplet size,
the system is again reduced to a 2D parameter space in imbalance and trap
frequency, $\left(\delta N_1, \omega_r\right)$. The remainder of this section
is split into two subsections: firstly, breathing modes are observed with
varying trap strengths with small imbalances, of a similar magnitude to those
in \cref{fig:gs_trap}(c); secondly, trap frequency is fixed allowing for
breathing modes to be observed with imbalances much larger than those in
\cref{fig:gs_trap}(c).

\subsection{Varying trap strength}

\begin{figure}
  \includegraphics[width=0.5\textwidth]{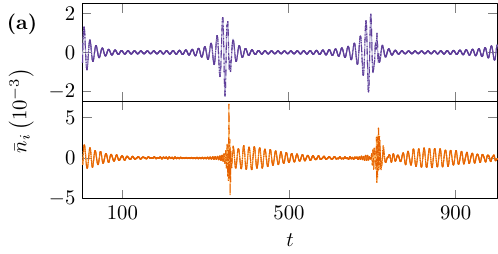}
  \includegraphics[width=0.5\textwidth]{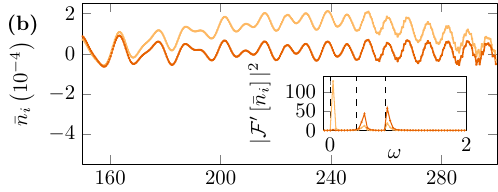}
  \includegraphics[width=0.5\textwidth]{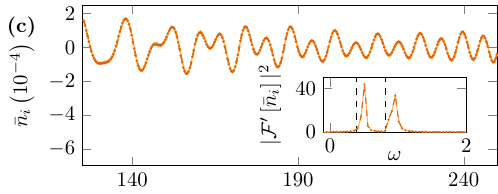}
  \includegraphics[width=0.5\textwidth]{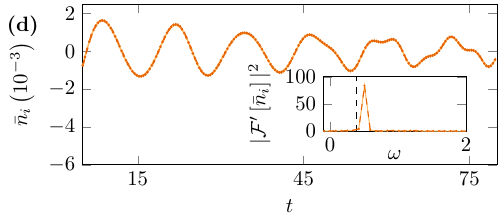}
  \caption{Breathing modes of trapped imbalanced droplets. (a) Droplet central
    densities in time for both a self-evaporative, balanced (upper) and
    imbalanced (lower) droplet (with imbalance $\delta N_1 \approx 4257$) in a
    trap with frequency $\omega_r \approx 0.00883$. The droplet is defined with
    the same parameters as in \cref{fig:gs_trap}. Note the recombination events
    that cause the self-evaporative dynamics to be reinitiated. (b) Droplet
    central density for the $\delta N_1 \approx 4257$ imbalanced droplet in the
    lower panel of (a), in a trap with $\omega_r \approx 0.00662$. The trap
    frequency is sufficiently low such that all three modes --- i.e., the
    intrinsic mode, and the two modes corresponding to the two chemical
    potentials --- can be observed before the first recombination event. (c) A
    higher trap frequency of $\omega_r \approx 0.0106$ showing that the shorter
    period between recombination events no longer allows for the long
    wavelength oscillation of the majority component. (d) An increased trap
    frequency of $\omega_r \approx 0.0309$. The recombination events occur
    within such short intervals that the dynamics are dominated by the
    intrinsic droplet breathing mode as there is not sufficient time for this
    mode to decay.}
  \label{fig:SE_trap}
\end{figure}

The upper panel of \cref{fig:SE_trap}(a) shows an example of a
self-evaporative, balanced droplet in a trap of frequency $\omega_r \approx
0.00883$. The data shown is a measure of the droplet central density,
$\bar{n}_i(t) = n_i(r=0,t) - \langle n_i(r=0)\rangle_t$, where
$\langle\cdots\rangle_t$ represents time averaging. The droplet exhibits a
decaying oscillation due to the emission of particles, causing the droplet to
asymptotically relax to a lower energy state. However, the emitted particles
are refocused by the trap back toward the droplet resulting in the short-lived,
high-amplitude oscillations, which are the result of a recombination event
between the droplet and the reabsorbed wavepacket. This then leads to the
self-evaporation reoccurring at set intervals of approximately half the
associated trap period of $T = 2\pi/\omega_r \approx 712 = 2\times 356$ with $t
= 356$ being the approximate time for the reinitialised decay in
\cref{fig:SE_trap}(a). 

The balanced droplet recombination can be thought of as `clean', as there is
little noise produced and the reinitialised oscillation is approximately
equivalent to the initial oscillations. In the presence of an imbalance
($\delta N_1 \approx 4257$), given in the lower panel of \cref{fig:SE_trap}(a),
the recombination events are not `clean' as each separate repetition of
decaying oscillation is not equivalent to the previous. The trapped, unbound
atoms alter the recombination of the emitted particles. Eventually these
recombination events will lead to significant noise and thus the remainder of
the analysis presented here will focus on the dynamics prior to the first
recombination event.

\cref{fig:SE_trap}(b), (c) and (d) show the same self-evaporative droplet given
in \cref{fig:SE_trap}(a) with an imbalance of $\delta N_1 \approx 4257$, for
three trap frequencies $\omega_r \approx \{0.00662,0.0106,0.0309\}$, with each
figure showing times before the first recombination event. The data presented
is the same measure of central density as in \cref{fig:SE_trap}(a), with insets
of the associated power spectra $|\mathcal{F}^{\prime}[\bar{n}_i]|^2$ in which
$\mathcal{F}^{\prime}[\cdot]$ denotes the power spectrum rescaled by the mean,
and all negative frequencies set to zero purely for better data visualisation.
The time periods are chosen to highlight the behaviour of the modes, largely
after the decay of the initial mode, except for in \cref{fig:SE_trap}(d),
discussed further below. 


The breathing mode dynamics of an imbalanced droplet in a trap of frequency
$\omega_r \approx 0.00662$ are given in \cref{fig:SE_trap}(b). These dynamics
exhibit the three distinct modes of the equivalent free-space droplet (with the
associated three peaks given in the inset power spectrum) \cite{flynn2022}. The
near-zero frequency peak in the power spectrum of the majority component
corresponds to the free-space majority component chemical potential, and is the
highest amplitude mode. The effect of this mode can be seen by the relative
difference in oscillation between the two components in \cref{fig:SE_trap}(b).
There is also the superposition of two other modes which are of comparable
amplitude in both the majority and minority component, corresponding to the
intrinsic droplet breathing mode (the central peak of the inset) --- i.e., the
initial, high amplitude mode --- and the minority component chemical potential
(the highest frequency mode in the inset).

Increasing the trap frequency to a balanced droplet corresponds to a relatively
small increase in breathing mode frequency \cite{hui2020}, and this effect
appears to carry over to the trapped, imbalanced droplet. The inset power
spectrum of \cref{fig:SE_trap}(b) includes the three free-space breathing modes
given by the vertical, dashed lines. All three of these modes have an upshifted
frequency due to the trap, though this increase is small due to the relatively
low trap frequency.

\cref{fig:SE_trap}(c) shows that if the trap frequency is increased, eventually
the highest amplitude mode is lost. By increasing the trap frequency to $\omega
\approx 0.0106$, the two component central densities oscillate in phase with
one another, i.e., there is no long-wavelength oscillation between the two
components as given in (b). The low frequency mode cannot oscillate within the
reduced period between recombination events from the increased trap frequency.

\cref{fig:SE_trap}(d) shows the highest trap frequency, $\omega_r \approx
0.0309$, considered in this section. At this trap frequency, the
period between recombination events is considerably shortened. Hence the time
window in focus is dominated by the initial, high-amplitude mode of the
droplet, shown by the single peak in the inset power spectrum. There are some
interactions with other modes at the later times shown in
\cref{fig:SE_trap}(d), but there is not sufficient time for the initial mode to
decay. 

No higher trap frequencies are studied here because the high recombination rate
implies only the intrinsic mode is observable. Note too that this highest
trapping frequency is still an order of magnitude smaller than $\omega_r^{(c)}
\approx 0.188$. 

In summary, there is a close relationship in the dynamics between imbalanced
droplets in free-space and in harmonic traps. For low trap frequencies the
three breathing modes of the free-space, imbalanced droplet are visible.
However, increasing the trap frequency leads to the loss of the majority
component mode. Eventually for higher trap frequencies, the oscillations are
dominated by the intrinsic droplet mode, as there is not sufficient time
between recombination events for the initial mode to decay, resulting in the
loss of the minority component mode. The recombination events in higher
frequency traps lead to dynamics rapidly dominated by excitations. Therefore,
if the multiple breathing modes of trapped imbalanced droplets were to be
experimentally observed, it is advised to use low trap frequencies, such that
the initial intrinsic mode can sufficiently decay.

\subsection{Varying Imbalance}

Having established how the breathing modes of imbalanced droplets vary with
trap frequency, this section focuses on how these modes vary with increasing
imbalance. To analyse the breathing modes as a function of imbalance, the
weakest trap strength studied in \cref{fig:SE_trap} is used, because all three
imbalanced droplet breathing modes are observable.

\begin{figure}
  \includegraphics[width=0.5\textwidth]{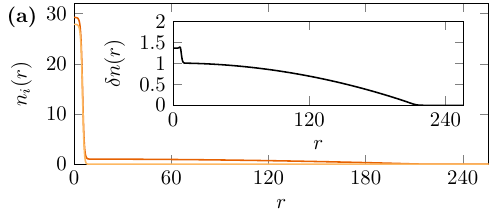}
  \includegraphics[width=0.5\textwidth]{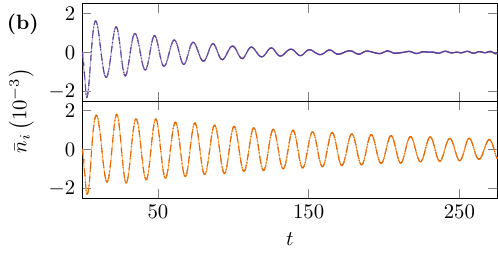}
  \includegraphics[width=0.5\textwidth]{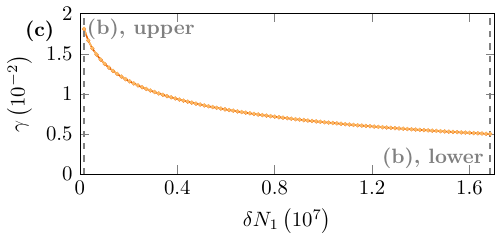}
  \caption{Ground states and breathing modes for high imbalances in a low
    frequency trap. (a) Example ground state density profile in a trap of
    frequency $\omega_r \approx 0.00662$, and an imbalance of $\delta N_1
    \approx 16856418$. (b) Two examples of breathing modes with a smaller
    imbalance of $\delta N_1 \approx 170267$ in the upper panel, and a
    considerably larger imbalance of $\delta N_1 \approx 16856418$ in the lower
    panel. These example imbalances are highlighted in (c) by the vertical
    dashed lines. (c) Fitted decay rate to the initial droplet breathing mode,
    with varying size of imbalance.}
  \label{fig:high_imb_sims}
\end{figure}

\cref{fig:high_imb_sims}(a) shows an example ground state density profile of a
weakly-trapped, highly imbalanced droplet with the majority and minority
components, shown in dark and light orange, respectively. The inset shows the
density difference, $\delta n(r) = n_1(r) - n_2(r)$, between the majority and
minority component. The density structure within the droplet core is comparable
to the small imbalances shown in \cref{fig:gs_trap}(a). However, the key
difference with highly imbalanced mixture is the large radius gas surrounding
the droplet.

\cref{fig:high_imb_sims}(b) highlights two examples of breathing mode
oscillations for droplets with high population imbalances. The upper panel
shows a mixture with $\delta N_1 \approx 170267$, whilst the lower panel shows
a mixture with $\delta N_1 \approx 16856418$. These population imbalances are
so large due to the weak trap geometry used, i.e., to achieve significant gas
densities in these low frequency traps, substantial imbalances are needed. The
same measure of central density as used in \cref{fig:SE_trap} is shown, prior
to the first recombination event.

In both panels of \cref{fig:high_imb_sims}(b) there does not appear to be any
out-of-phase oscillations between the two components from the
majority-component chemical potential mode, as shown in \cref{fig:SE_trap}(b).
The surrounding gas therefore seems to have frozen out this long-wavelength
mode, similar to the tighter trap in \cref{fig:SE_trap}(c). The main difference
between the upper and lower panels of \cref{fig:high_imb_sims}(b) is the lower
decay rate of the initial, high-amplitude mode in the more imbalanced mixture.
As shown in \cref{fig:SE_trap}(b), at later times the initial mode in the upper
panel has decayed sufficiently such that the minority-component chemical
potential mode is visible. By driving the imbalance even higher the decay rate
of the initial rate is greatly reduced, such that minority-component chemical
mode cannot be observed.

In \cref{fig:high_imb_sims}(c), the droplet central density is fitted to a
decaying sinusoidal curve, of the form $\bar{n}_i = Ae^{-\gamma t}\sin(\omega t
+ \phi) + c$, using the optimize.curve\_fit function from the SciPy library for
Python \cite{2020scipy}. With increasing imbalance there is a corresponding
decrease in the fitted decay rate, $\gamma$. This implies that a higher
surrounding gas density resists the particle emission from the droplet. This
could be of potential benefit to experiments as it implies that larger
imbalances would give more time to observe the decaying breathing mode
oscillations.

Beyond the observation of trapped imbalanced droplets, there are also questions
to be asked of the experimental realisability of a free-space imbalanced
droplet. For example, how stable is the imbalance under the transition from the
trap --- used to prepare the mixture --- to the free-space droplet.

\section{RELEASE INTO FREE SPACE \label{sec:tof}}

This section studies how imbalanced droplets behave when the trap is turned
off, releasing the droplet into free space. This is similar to a Time-Of-Flight
(TOF) expansion, a method used in experiments in which trap potentials are
switched off and the atomic cloud expands, often used for imaging. Imbalanced
droplets are less stably bound than balanced droplets \cite{flynn2022}, and
thus the motivating question is whether it is possible to preserve the
imbalance when released from a trap. This is a crucial question in the
feasibility of experimentally creating a free-space, imbalanced droplet.

TOF expansion is a widely used technique in quantum gas experiments
\cite{castin1996} from the very first experimental observation of Bose-Einstein
condensation \cite{anderson1995,ketterle1995}. Typically, by removing the trap
the resulting expansion increases the scale of defects, such as vortices,
accounting for the low resolution of imaging apparatus
\cite{madison2000,raman2001,abo2001}. Measurements of condensate density from
TOF images can also be used to compute approximate temperatures and population
numbers of the cloud \cite{ensher1996,stamper1998}.

Whilst most quantum gas experiments are inherently in the gas phase, droplets
are by definition self-bound, liquid states \cite{petrov2015}, and hence must
retain an approximately fixed size when released into free space. This property
has proved popular for experiments as evidence for the production of quantum
droplets
\cite{cabrera2018,semeghini2018,cheiney2018,derrico2019,guo2021,cavicchioli2022},
though relatively high-resolution imaging is necessary. 

The two observables used here to measure the dynamics resulting from the
release into free-space, are: the population numbers contained within the
droplet, and the central droplet density difference, $\delta n(r=0) = n_1(r=0)
- n_2(r=0)$. The population numbers are used to measure the particle loss from
each component, while the central density difference is used as a measure of
how the droplet core evolves after being released from the trap. The population
numbers of the droplet are computed by
\begin{equation*}
  N_i^{\text{drop}}(t) = 4\pi\int_{0}^{R^\text{drop}(t)}r^2|\Psi_i(r,t)|^2\dd r 
\end{equation*}
in which $R^{\text{drop}}(t)$ is defined as the radius at which the component
density equals $0.1\%$ of the maximum component density, i.e., giving an
approximate droplet radius. The population numbers are extracted in time, and
$R^{\text{drop}}(t)$ is allowed to vary dynamically. To simulate the release
into free space, ground states are computed as in \cref{sec:drop_gs_trapped},
the traps are then instantaneously turned off and the mixture is evolved in
real time. The instant trap turn off can be quite a violent excitation of the
droplet particularly with higher trap frequencies.

\begin{figure*}
  \begin{tabular}{cc}
    \includegraphics[width=0.5\textwidth]{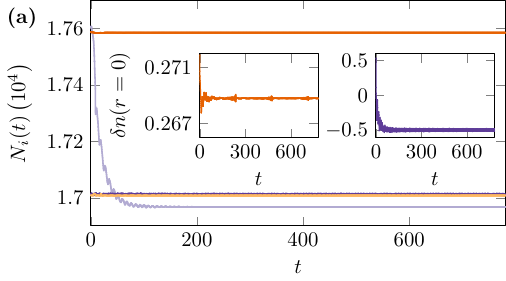} &
    \multirow{2}{*}[\dimexpr0.27\textwidth]{{ \includegraphics[width=0.5\textwidth]{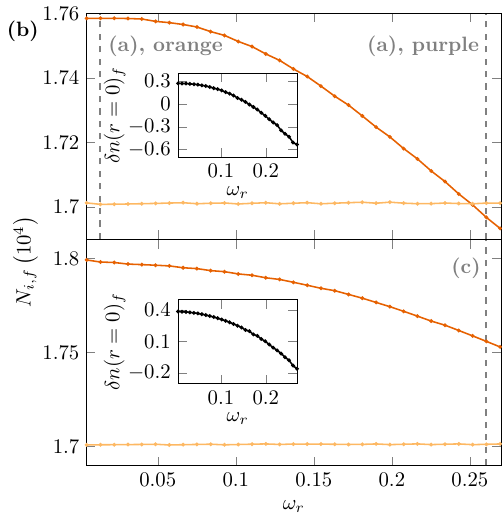}}} \\
    \includegraphics[width=0.5\textwidth]{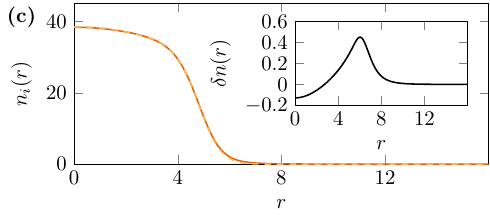}
  \end{tabular}
  \caption{Dynamics of imbalanced droplets, after release into free space, with
    the same parameters as in Figures \ref{fig:gs_trap}, \ref{fig:SE_trap} and
    \ref{fig:high_imb_sims}. (a) Population numbers, with an imbalance of
    $\delta N_1 \approx 596$, from dynamics resulting from a release into free
    space with orange colours corresponding to an initial low frequency trap,
    with $\omega_r\approx 0.00442$ (dark and light orange corresponding to
    majority and minority components, respectively), and purple colours
    corresponding to an initial tight trap, with $\omega_r \approx 0.269$
    (light and dark purple corresponding to majority and minority components,
    respectively). The insets show the differences in central densities through
    time, with the left and right panels corresponding to the low and high
    frequency trap cases, respectively. These two example simulations are
    highlighted in the upper panel of (b), by the vertical, grey, dashed lines.
    (b) The late-time population numbers and central densities difference
    (inset) for varying initial trap frequencies, in the range $0.00442\lesssim
    \omega_r \lesssim 0.269$. The upper panel corresponds to a bound,
    imbalanced droplet ($\delta N_1 \approx 596$) whilst the lower panel
    corresponds to a saturated, imbalanced droplet with an external unbound gas
    ($\delta N_1 \approx4257$). (c) A real time density profile example (with
    an inset of the density difference) from the late time free-space release
    dynamics (with initial imbalance $\delta N_1\approx 4257$ and high trap
    frequency $\omega_r \approx 0.269$) [corresponding to the right-hand,
    vertical line in the lower panel of (b)]. The light and dark orange curves
    correspond to the minority and majority components, respectively. Note the
    negative central density difference showing the reversal of the imbalance
  within the droplet core.} 
  \label{fig:tof_sims}
\end{figure*}

\cref{fig:tof_sims}(a) shows the two-component population numbers for two
different initial trap frequencies: a lower frequency of
$\omega_r\approx0.00442$, given by the orange curves, and a higher frequency of
$\omega_r\approx0.269$, given by the purple curves. The droplet size is the
same as in Sections \ref{sec:drop_gs_trapped} and \ref{sec:breath_trap}, and is
within the bound, imbalanced regime ($\delta N_1 \approx 596$), i.e., there is
no surrounding gas. The population numbers of both the minority (light orange)
and majority (dark orange) components, in the lower frequency trap, remain
relatively constant, as does the central density difference given in the left
inset. There are some small oscillations present in the central density
difference, which appear to be excitations from the instantaneous release into
free space. 

The droplet in the higher frequency trap has a relatively constant
minority-component (dark purple) population number, whereas the majority component
(light purple) undergoes significant losses. The losses are so high that the
imbalance of the droplet is reversed, as can be seen by the sign reversal of
$\delta n(r=0)$. 

Following the initial transient stage of heavy majority component losses, the
droplet equilibrates with some long-lived small oscillations. Therefore,
\cref{fig:tof_sims}(a) demonstrates that the equilibrated imbalance following
the release into free space, depends on the initial trap frequency.
\cref{fig:tof_sims}(b) shows the trap frequency dependency of the late-time
equilibrated population numbers, $N_{i,f}$. Presented are the majority (dark
orange) and minority component (light orange) of a bound, imbalanced droplet
(upper panel) and a saturated, imbalanced droplet surrounded by an unbound gas
(lower panel), with the late-time central density difference, $\delta
n(r=0)_f$, inset. 

The upper panel of \cref{fig:tof_sims}(b) demonstrates that the imbalance can
be approximately conserved following the release from an initial low frequency
trap. However, this imbalance is lost or even partially reversed with higher
trap frequencies. A larger imbalance of $\delta N_1 \approx 4257$ is shown in
the lower panel. The main difference between these two panels is that higher
initial imbalances suppress the imbalance reversal to higher initial trap
frequencies. 

Following the release the into free space, the density profile of an
equilibrated, imbalanced droplet can starkly differ from an identically
imbalanced ground state droplet. The reversal of the central density difference
occurs at smaller values of $\omega_r$ than the reversal of the population
numbers. This suggests that these equilibrated droplets are stably excited
states that exhibit a higher minority component density in the droplet core. An
example of this is shown in the real-time density profiles of an initially
saturated droplet, with $\delta N_1 \approx 4257$, shown in
\cref{fig:tof_sims}(c) [corresponding to the grey, vertical line in the lower
panel of (b)]. This real-time snapshot shows that the droplet can stabilise to
$\delta n(r) < 0$ in the droplet core, with a small region of $\delta n(r) > 0$
at the droplet surface, i.e., some majority component density can stably reside
on the droplet surface. 

To summarise, an imbalanced droplet prepared in a relatively low frequency trap
can retain the majority of the initial imbalance after the release into free
space. However for increasing trap frequency, the initial majority component
begins to lose atoms until the imbalance is either negligible or reversed. The
core density reversal can occur even if the original majority component still
contains more atoms. This results in a stable, excited state in which some of
the majority component atoms sit at the droplet surface. The imbalance reversal
can be suppressed by preparing the droplet with a higher imbalance, i.e., a
higher density surrounding gas. These results show that imbalanced droplets can
be robust to a release into free space, suggesting that free-space,
imbalanced droplets are feasible using modern experimental techniques.

\section{DISCUSSION \label{sec:disc}}

This work has investigated ground states, breathing modes and the release
into free space of imbalanced droplets confined in isotropic harmonic traps.
First, \cref{sec:drop_gs_trapped} demonstrates that the trapping potential
squeezes any unbound gas up to the droplet, forming a significant gas density
at the droplet surface. The imbalance dependent divergence in the majority and
minority component chemical potentials, increases further with higher trap
frequencies.

\cref{sec:breath_trap} focused on the breathing modes of imbalanced droplets,
and contrasted the trapped geometry with the free-space results of
Ref.~\cite{flynn2022}. This section highlighted that the presence of a trap
causes recombination events from reflected particle. For a free-space
imbalanced droplet there are three breathing mode frequencies \cite{flynn2022}.
These three modes can be observed in the trapped, imbalanced droplet though
\cref{sec:breath_trap} shows that with increasing trap frequency these modes
are lost. Similarly the decay rate of the imbalanced droplet breathing mode can
be reduced by the presence of a significant majority-component gas.

The final results presented are the dynamics from releasing the imbalanced
droplets into freespace given in \cref{sec:tof}. The results show primarily
that with a low frequency initial trap, the droplet imbalance can be preserved
under trap release, however higher trap frequencies lead to a loss or inversion
of the imbalance. This gives promise for the experimental realisation of
free-space, imbalanced quantum droplets. 

The stability of the imbalance under release from a trap may be significant in
the experimental results of Refs.~\cite{semeghini2018, ferioli2019}, in which
the mixture is prepared with $N_2/N_1 = 1 \neq \sqrt{\aoo/\att}$. These works
assume that the droplet will dynamically balance, after the release into free
space, to $N_2/N_1 = \sqrt{\aoo/\att}$. This could explain why the data points
of $N_1/N_2$ in Fig.4(c) of Ref.~\cite{semeghini2018} are upshifted from the
balanced line of $N_2/N_1=\sqrt{\aoo/\att}$, as component-1 is setup to be the
majority component. Likewise, some of the results in Ref.~\cite{cheiney2018}
are speculated to be sensitive to imbalance. This work suggests that balanced
droplets could be a special case, and that imbalanced droplets are more common.

The analysis of spherically symmetric ground states and breathing mode dynamics
presented could be extended to explore heteronuclear mixtures. The different
kinetic energy contributions of the two components may lead to novel physics,
as adding an imbalance to either component is no longer symmetric. This is
however a non-trivial extension due to the form of the two-component LHY
correction of a heteronuclear mixture \cite{petrov2015,ancilotto2018}. 

The recombination events from the trap, limit the time for observing collective
modes. This restriction implies that smaller computational boxes could be used
to probe collective modes in more general 3D simulations, allowing for
observation of non-zero angular momentum modes such as dipole
\cite{capellaro2018,cavicchioli2022} and quadrupole modes in both balanced and
imbalanced droplets. 

The potential of new mixtures for probing droplet physics is exciting but some
experiments use highly anisotropic trap potentials with significant population
number imbalances \cite{wilson2020}. Therefore it vital to understand how
droplets are modified from the prototypical balanced, free-space profile, by
population imbalances and trap potentials.

The data presented in this paper are available \cite{taf_data}.

\begin{acknowledgments}
The authors acknowledge support from the UK Engineering and Physical Sciences
Research Council (Grants No. EP/T015241/1, and No. EP/T01573X/1). T. A. F. also
acknowledges support from the UK Engineering and Physical Sciences Research
Council (Grant No. EP/T517914/1). This research made use of the Rocket High
Performance Computing service at Newcastle University.
\end{acknowledgments}

\bibliography{references.bib}

\end{document}